\documentclass[usegraphicx,onecolumn, referee]{mn2e}
\usepackage{txfonts}

\def\beq{\begin{equation}}
\def\eeq{\end{equation}}
\def\kms{\,\rm km\,{s}^{-1}}

\def\mpc{\,\rm Mpc}
\def\LCDM{\Lambda{\rm CDM}}
\newcommand{\alphabf}{{\mbox{\boldmath$\alpha$}}}
\newcommand{\etabf}{{\mbox{\boldmath$\eta$}}}
\newcommand{\xibf}{{\mbox{\boldmath$\xi$}}}
\newcommand{\Ds}{D_{\rm s}}

\newcommand{\reference}{\bibitem}
\def\mnras{MNRAS}

\def\aap{A\&A}
\def\apj{ApJ}
\def\apjs{ApJS}
\def\aj{AJ}
\def\nat{Nature}

\title[Rotation in Gravitational Lenses]{Rotation in Gravitational Lenses}

\author[Pen \& Mao]{
Ue-Li Pen$^{1,2}$, Shude Mao$^{3}$ \thanks{upen@cita.utoronto.ca;
    smao@jb.man.ac.uk} \\
$^{1}$ National Astronomical Observatories, 
Chinese Academy of Sciences, A20 Datun Road, 100012 Beijing, China \\
$^{2}$ Canadian Institute for Theoretical Astrophysics, 60
St. George St., Toronto, M5S 3H8, Canada \\
$^{3}$ University of Manchester, Jodrell Bank Observatory, 
  Macclesfield, Cheshire SK11 9DL, UK
}

\date{
Accepted ........
Received .......;
in original form ......}

\pubyear{2004}

\begin{document}
\maketitle

\begin{abstract}
Gravitational lensing deflects light.  A single lens deflector can
only shear images, but cannot induce rotations.  Multiple lens planes
can induce rotations.  Such rotations can be observed in quadruply
imaged sources, and can  be used to distinguish between 
two proposed solutions of
the flux anomaly problem:  substructures in lensing galaxies vs large scale
structure.  We predict the expected amount of rotation due to large
scale structure in strong lensing systems, and show how this effect
can be measured using $\sim$mas VLBI astrometry of quadruple lenses with
extended source structures.
The magnitude of rotation is around one degree.  The biggest theoretical
uncertainty is the power spectrum of dark matter on very small scales.
This procedure can potentially be turned around to measure the dark
matter power spectrum on very small scales.
We list the predicted RMS rotation angles for several quadruple
lenses with known lens and source redshifts.
\end{abstract}


\section{INTRODUCTION}

Gravitational lensing results from the deflection of light under the
gravitational influence of all matter, luminous or otherwise. Its
physics is clean, and this effect has allowed the measurement of
the distribution of dark matter from galaxy-scales (using
strong lensing) to the large-scale structure of the universe (using weak
lensing). 

Usually, many approximations are made to simplify the calculations.
One of these is the Born approximation, where one calculates a small
deflection along the unperturbed light path.  Some effects, such as
image rotation due to multi-plane lensing, are not accessible in this
approximation.  Authors have obtained different results for the
magnitude of multi-plane weak lensing rotation (Jain et al 2000, 
Cooray \& Hu 2002, hereafter CH;
Hirata \& Seljak 2003, hereafter HS). 
Schneider (1997) showed that
a strong lens plus spatially constant 
weak lens system is mathematically identical
to some other single lens plane system, so only differential rotation
at the image positions is observable.

In this paper, we apply the multiple plane lensing calculation to real
physical systems: quadruply imaged quasars.  We show in this
physical example how the rotation effect can be measured, why it is physical
and real, estimate its magnitude, and show how it can be used to
resolve the substructure controversy.

The rotation of images is of current interests in the context of
using gravitational lensing to detect substructures predicted by
the Cold Dark Matter (CDM) structure formation model.  From 
both semi-analytical studies and numerical simulations,
it became clear that hundreds of subhaloes (substructures) are 
predicted to exist in a Milky-Way type halo
(e.g., Kauffmann et al. 1993; Klypin et al. 1999;
Moore et al. 1999; Ghigna et al. 2000). In general, about 5-10\% of 
the mass is predicted to be in substructures, with a typical
mass spectrum of $n(M)dM \sim M^{-1.8}dM$. If all the 
substructures form stars, then the predicted number of
satellite galaxies exceeds the observed number
in a Milky-Way type halo by a large factor. It is now, however, clear
that the correspondence between substructures and visible
satellite galaxies is not simple (Gao et al. 2004a,b; see also Springel et
al. 2001; Diemand, Moore \& Stadel 2004; Nagi \& Kravtsov 2005).
In particular, if only some substructures house satellite galaxies,
then the discrepancy can
be alleviated (e.g., Kravtsov et al. 2004). At present, it is
not entirely clear whether the internal kinematics of satellite galaxies
are consistent with observations (Stoehr et al. 2002; Kazantzidis et
al. 2004). Furthermore, the spatial distribution of satellite galaxies 
in the Milky Way is also somewhat puzzling (Kroupa et al. 2004, but see
Kang et al. 2005; Liebeskind et al. 2005; Zentner et al. 2005).

One possible way to detect the (dark) substructures is through
the gravitational lensing effect. Simple analytical models
in gravitational lenses often fail to reproduce
the observed flux ratios (e.g., Kochanek 1991).
This discrepancy is commonly  referred to as the ``anomalous flux ratio problem.''
This has been proposed as evidence for substructures in the primary
lensing galaxies (e.g., Mao \& Schneider 1998;
Metcalf \& Zhao 2002; Dalal \& Kochanek 2002). However, as most of the
predicted substructures are in the outer part of the lensing galaxies
while the lensed images are typically at a projected distance of only
a few kpc from the center, it is 
unclear whether the predicted amount of substructures in lensing galaxies
by CDM is sufficient (Kochanek \& Dalal 2004;
Mao et al. 2004), so it is important to consider other sources of ``substructures.''
Recently, Metcalf (2005) proposed that substructures along the line
of sight can equally explain the discrepancy. A key question naturally
arises: how do we know that substructures are from the primary
lens or from elsewhere along the line of sight?

In this paper, we examine the rotation of images induced by structures
along the line of sight in gravitational lenses (see also Chen et
al. 2003).  We use a novel power-spectrum approach and consider the
fluctuations of surface densities among different images (separated by
few tenths to few arc seconds) and their effects on the magnifications.
Throughout this paper, we adopt the ``concordance'' $\LCDM$ cosmology
(e.g., Ostriker \& Steinhardt 1995; Spergel et al. 2003 and references
therein), with a density parameter $\Omega_{\rm m}=0.3$, a cosmological
constant $\Omega_{\Lambda}=0.7$, a baryon density parameter
$\Omega_{\rm b}=0.024h^{-2}$, and the power-spectrum
normalization $\sigma_8=0.9$. We adopt a  Hubble constant of
$H_0=70\kms\mpc^{-1}$.

\section{Lensing by Large Scale Structure in the presence of a strong lens}

Many multiply-imaged gravitational lenses on galaxy-scales have been observed. At the
time of writing, roughly 100 such systems
are known\footnote{see the CaSTLES database:
  http://cfa-www.harvard.edu/castles/}.  The largest systematic survey,
the Cosmic Lens All Sky Survey (CLASS) has found 22 new galaxy-scale
lenses, approximately one half of which are quadruple lenses (Browne et al. 2003;
Myers et al. 2003). Most of these have high-resolution imaging from
$0.1\arcsec$ to mas from MERLIN, HST to VLBI. Several of these lenses
are resolved into multiple components. The system 0128+437
provides a good example (Biggs et al. 2004). Each of the image has been resolved into
three sub-components with VLBI. Such high-resolution images
provide an excellent test bed for lensing models.

In addition to the strong lens system which causes the multiple image
splitting, all the matter along the line of sight will further deflect
the light and contribute to distortions in the image.
One such effect is the apparent rotation of images.  It can be shown
that a single plane lens has a symmetric amplification matrix, which
shears but does not rotate images.  With multiple planes, image rotation
is possible.  Observing it may appear non-trivial, since it would require prior
knowledge of the unlensed image alignment.  We will show below (see \S3) how this
can actually be measured if one has a quadruply imaged source with
extended structures.

Several geometric configurations can lead to image rotation.  A
strong lens has shear and convergence of order unity.  With
sufficiently accurate alignment, a second strong lens could occur
along the line of sight. Indeed such an example has already been
seen -- the JVAS/CLASS lens system, B2114+022 (Augusto et
al. 2001; Chae, Mao \& Augusto 2001), has two lensing galaxies at redshifts
0.3157 and 0.5883 respectively, within 2 arc seconds of 
quadruple radio sources\footnote{It is present unclear whether 
the radio sources B and C are lensed images.}.

But in general, the expected variation in surface density due to dark
matter is small.  Integrating the Limber equation, this leads to large
scale structure density variations of order a few percent of the
critical surface density (e.g., Jain et al 2000). 
So a random cluster lens will typically only have weak lenses in its
foreground and background.

The typical splitting angle of the strong lens is around one arc
second.  Large scale structure
density fluctuations on such scales are significantly correlated.
Since Schneider (1997) had shown that perfectly correlated weak
lensing screens do not cause observable rotation, one must compute
the differential weak lensing shear.

\begin{figure}
\centering\includegraphics[height=9cm]{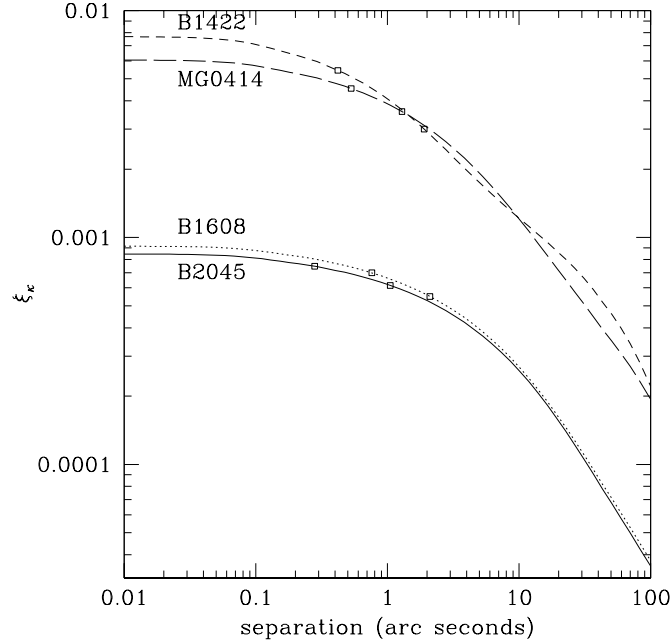}
\caption{The correlation function for convergence ($\kappa$)
for different image pairs induced by the large-scale structure; the
correlation function at zero lag is the variance in $\kappa$.
The boxes indicate the largest and smallest splittings for four CLASS lenses
with known lens and source redshifts. Because splitting
angles are small, the $\kappa$ by the large-scale structure at different image positions is
strongly correlated.  The variance from large scale structure shear at
each image position is the correlation function at zero lag.  The
covariance for images separated by a lag $\Delta \theta$ can be seen
in the plots. 
Since the covariance (at typical image separations)
is of comparable size of to 
the auto-variance, which leads to a supression of the differential
shear mode.
}
\label{fig:corr}
\end{figure}

The cross correlation between two image positions
separated by angle $\Delta \theta$ is defined as 
\begin{equation}
r \equiv \xi_\kappa(\Delta \theta)/\xi(0),
\end{equation}
where $\xi_\kappa(\Delta\theta)$ is the two dimensional
Fourier (Bessel) transform of Equation (\ref{eqn:limber}) which
will be discussed below. The results are shown for 
four quadruple lenses with known lens and source redshifts,
B1422+231 (Patnaik et al. 1992), 
MG0414+0534 (Hewitt et al. 1992), B1608+656 (Myers et al. 1995), and
B2045+265 (Fassnacht et al. 1999).
We find a ratio of variances between the 
difference of two images $\sigma^2_-$ and the individual variances to be
\beq
\frac{\sigma^2_-}{\sigma^2}=2(1-r).
\eeq
If the two images are uncorrelated, the difference will have twice the
variances of each individual image.
Figure \ref{fig:corr} shows the correlation function vs. image
separation. As can be seen, typically we have
$r>0.5$, so the difference mode has a slightly lower variance than each
individual image.

\begin{figure}
\centering\includegraphics[height=9cm]{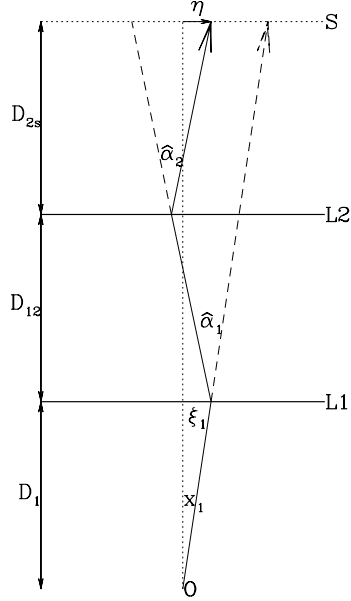}
\caption{Lensing geometry.  The deflection angles themselves are not
 observable,  but rather the changes in deflection angle. Various
quantities are indicated in the Figure and used in eqs. (\ref{eqn:x}-\ref{eqn:lens}).
}
\label{fig:geometry}
\end{figure}

We assume that the various
images pass through different parts of the strong lens with correspondingly
different values of shear and convergence, and
also differing weak lens deflectors.  
We consider a 2-plane
lens, $L_1$ is the strong lens, $L_2$ is a weak lens due to large
scale structure, $O$ is the observer position and $S$ is the source
plane.  The geometry is shown in Figure \ref{fig:geometry}.
We consider a quadratic potential on each lens plane,
$\phi=ax_1^2+2bx_1 x_2+c x_2^2$, where $(x_1, x_2)$ are the
(angular) coordinates in the lens plane. 
The units for the potential are chosen such
that $2\kappa=\bigtriangledown^2 \phi$,
and the deflection angle is $\hat{\alphabf}=\bigtriangledown_{x} \phi$ due to
lens, where we use the same notations as in Schneider, Ehlers, \& Falco (1992).
To distinguish the two lens planes, we will use a prime to
denote variables in the $L_2$ plane.  

For a general anisotropic lens, the deflection angle is a vector. For
our quadratic potential, $\kappa$ and $\gamma_{1,2}$ are constant on the plane,
\beq
\kappa={1 \over 2}(a+c), ~~\gamma_1={1\over 2}(a-c), ~~ \gamma_2=b.
\eeq
The quadratic potential is the most general function which leads to
constant values of $\kappa$ and $\gamma$.  Their constancy on small
scales is inferred from the correlation function shown in Figure
\ref{fig:corr}.
In such a potential, the full gravitational lensing effect is straightforward
to compute.  One can solve the full photon trajectory, which are deflected
on the lens planes by the gradient of the potential.

We now denote $D_{1}$ to be the angular diameter distance from the
observer to the first lens, $D_{12}$ the distance from the first to
the second lens, and $D_{\rm 2s}$ the distance from the second lens to the
source, and so on.
We see from Figure \ref{fig:corr} that all the relative
deflection angles 
are very small in units of radians ($\sim 10^{-5}$), so we can expand to
first order in the deflection angle (Schneider et al. 1992, chapter 9)
\beq
{\etabf}={\Ds \over D_1} {\xibf}_1 -D_{\rm 1s}\hat{\alphabf}_1({\xibf}_1)
-D_{\rm 2s} \hat{\mathbf \alphabf}_2({\xibf}_2), ~~ {\xibf}_2={D_2 \over D_1} {\xibf}_1 - D_{12} \,\hat{\alphabf}_1({\xibf}_1).
\label{eqn:x}
\eeq
where $\mathbf{\etabf}$, ${\xibf}_1$, ${\xibf}_2$ denote the position vectors in the source plane
and the first and second lens planes (see Fig. \ref{fig:geometry}), and
the deflection angles in the first and second planes are given as
$\hat{\alphabf}_1$ and $\hat{\alphabf}_2$.

Following Schneider et al. (1992), we define two reduced deflection angles
\beq
{\alphabf}_1 = {D_{\rm 1s} \over D_s} {\bf \hat \alphabf}_1,~~
{\alphabf}_2 = {D_{\rm 2s} \over D_s} {\bf \hat \alphabf}_2,
\eeq
With these, eq. (\ref{eqn:x}) can be recast in a very simple form using
only angles:
\beq
{\bf y}={\bf x}_1 - {\alphabf}_1({\bf x}_1) - {\alphabf}_2({\bf x}_2),
~~
{\bf x}_2 = {\bf x}_1 - \beta_{12} {\alphabf}_1({\bf x}_1),
\eeq
where  ${\bf y}=\etabf/D_s$, ${\bf x}_1={\xibf}_1/D_1$, ${\bf x}_2={\xibf}_2/D_2$, and
$\beta_{12}= {D_{12} D_s/(D_2 D_{\rm 1s})}$.

Lensing shear and magnification can be obtained by studying the change of source position
${\bf y}$ resulting from a change of apparent angular position ${\bf x}_1$. In
particular, the magnification is given by (Schneider et al. 1992)
\beq
{\bf A} = {\partial{\bf y} \over \partial {\bf x}_1}
={\cal I}  - {\bf U}_1 - {\bf U}_2 + \beta_{12} {\bf U}_2 {\bf U}_1,
\label{eqn:lens}
\eeq
where ${\cal I}$ is a unit matrix,
$
{\bf U}_1 = \partial{{\alphabf}_1}/\partial{{\bf x}_1}$, and
${\bf U}_2 = \partial{{\alphabf}_2}/\partial{{\bf x}_2}$.

We are considering the combined effects of a strong and a weak lens.
The contribution of the weak lens ${\bf U}_2$ is small and we are
only interested in its contribution to rotation, so we neglect
its linear effect. In the linear (i.e. Born approximation) regime, it
is also observationally 
not possible to distinguish between contributions from the strong and
weak lens.
In the product term, we can absorb $\beta_{12}$
into the definition of $U_2$.  We define the critical density for the
weak lensing large scale structure to be 
\beq
\Sigma^{\rm crit}_2\equiv \frac{c^2}{4\pi
  G}\frac{D_{\rm 1s}}{D_{12}D_{\rm 2s}},
\label{eqn:sigma}
\eeq
which is the lensing strength of the large scale structure as seen by
an observer at the strong lens position.  This lends a simple
observational and computational interpretation of lensing rotation:
all distortions visible to an observer at the strong lens position
enter linearly into the coupling product of strong and weak lens.

An analogous results holds when the weak lensing plane is in front of
the strong lens.  In our notation, we have 
\beq
{\bf U}_1=\left(\begin{array}{cc} \kappa+\gamma_1 & \gamma_2 \\ \gamma_2 & \kappa -\gamma_1 \end{array}\right), \ \ 
{\bf U}_2=\left(\begin{array}{cc} \kappa'+\gamma_1'&\gamma_2'  \\  \gamma_2'&\kappa'-\gamma_1' \end{array}\right)
\label{eqn:a12}
\eeq
where $\kappa'\equiv \Sigma_2/\Sigma^{\rm crit}_2$,
and so
\begin{eqnarray}
{\bf A} &= & \left(\begin{array}{cc} \gamma_2\gamma_2'
+(1-\kappa+\gamma_1)(1-\kappa'+\gamma_1') &\gamma_2'(1-\kappa-\gamma_1)
+\gamma_2(1-\kappa'+\gamma_1') \\  
\gamma_2'(1-\kappa+\gamma_1)+\gamma_2(1-\kappa'-\gamma_1')&\gamma_2\gamma_2'
+(1-\kappa-\gamma_1)(1-\kappa'-\gamma_1') \end{array}\right) 
\nonumber \\
&\sim& \left(\begin{array}{cc} 1-\kappa+\gamma_1 &\gamma_2+\gamma_2\gamma_1' 
\\  \gamma_2-\gamma_2\gamma_1'&1-\kappa-\gamma_1 \end{array}\right).
\label{eqn:a}
\end{eqnarray}
In the last approximate equality we used the limit that the weak lensing plane
(the primed variables) are much smaller than the strong lens plane,
but keeping the anti-symmetric piece which is relevant for rotations.
We define the off-diagonal antisymmetric component as $\omega \equiv \gamma_2 
\gamma_1'$.

For strong lens systems, the shear on $L_1$ is of order unity,
while the large-structure shear ($\gamma'_{1}$, $\gamma'_2$) is
a few percent, corresponding to a rotation of order a degree.
Rotation only results when the principal axes of the two amplification
matrices are misaligned.  We choose the coordinates such that
\beq
{\bf U}_1=\left(\begin{array}{cc} \gamma & 0 \\ 
                                       0 & -\gamma \end{array}\right), \ \ 
{\bf U}_2=\left(\begin{array}{cc} \gamma_1'&\gamma_2'  \\  
                                  \gamma_2'&-\gamma_1' \end{array}\right).
\eeq
$\gamma_1'$ does not contribute to rotation, so we set it to zero.
Their product is a pure antisymmetric matrix
\beq
{\bf U}_2 {\bf U}_1=\left(\begin{array}{cc} 0&\gamma \gamma_2'  \\  
                             -\gamma\gamma_2'&0 \end{array}\right). 
\label{eqn:asym}
\eeq
When added to a unit matrix, this corresponds to a rotation matrix by an
angle $\gamma \gamma_2'$.  

In equation (\ref{eqn:a}), we factor the amplification
matrix ${\bf A}={\bf A}_{\rm s}
{\bf R}(\phi)$, as a product of a symmetric matrix and a pure
rotation, where the rotation matrix  
\beq
{\bf R}=\left(\begin{array}{cc} \cos\phi&\sin \phi  
			\\  -\sin \phi &\cos\phi \end{array}\right).
\eeq
We have
\beq
{\bf A}=\left(\begin{array}{cc} 1-\kappa+\gamma_1 &\gamma_2+\omega \\  \gamma_2-\omega&1-\kappa-\gamma_1 \end{array}\right),
\eeq
where $\omega$ indicates the importance of rotation.
Then we find 
\beq
\tan\phi=-\frac{\omega}{1-\kappa}.
\label{eqn:sinphi}
\eeq
The non-rotating amplification matrix ${\bf A}_s$ has the same
convergence as ${\bf A}$ up to ${\cal O}(\omega^2)$, 
but has its shear components rotated by $\phi/2$.

The different images pass through different parts of the strong lens,
each with its own values of shear $\gamma$.  The rotation angle will
thus be different for each image.
The actual value of the large scale structure shear depends on the
non-linear power spectrum of dark matter at small physical scales.
This is both a function of the primordial power spectrum and its
slope, and non-linear gravitational physics.  The most accurate models
seem to be a 
combination of N-body simulations and heuristic models based on stable
clustering.  Calibrations at larger scales indicate consistency at the
better than 20\% level on length scales larger than about 100 kpc.  In
the standard models, the contribution from smaller scales is not
dominant, so one would expect forecasts to be good to a factor of two
(Huffenberger \& Seljak 2003).

To forecast the expected RMS rotation angle, we compute
the expected differential variation in the weak lens screen shear, and apply to
eqs. (\ref{eqn:asym},\ref{eqn:sinphi}).  Mathematically, shear is
a polarization field, which can be described as a trace-free spin-2
tensor field.  In two spatial dimensions, any trace-free spin-2 tensor
field can be decomposed into two kinematic scalars: the
``divergence-like'' component, also known as ``E''-mode which is
longitudinal to its Fourier decomposed wave vector, and a
pseudo-scalar ``curl-like'' or ``B''-mode which is transverse to its
wave vector.  In weak gravitational lensing, the ``E''-mode is
identical to the convergence field $\kappa$, and the ``B''-mode is
zero.  To compute the statistics of the shear, it thus suffices to
calculate the dimensionless variations in projected matter surface
density.  Its  variation is given by the Limber equation.  
It involves the projection of a three dimensional non-linear
power spectrum 
\begin{equation}
\Delta^2(k,z)\equiv\frac{k^3}{2\pi^2}P(k,z)
\end{equation}
to a two dimensional angular power spectrum $l(l+1)C_l/2\pi$
(Limber 1954; Kaiser 1992, 1998),
\begin{equation}
\frac{l(l+1)}{2\pi} C_l=\frac{\pi}{l}\int_{z_i}^{z_f}
\Delta^2(l/\Delta\chi(z),z) \,
w(z)^2 \, \chi(z) \,
\frac{d\chi}{dz} \, dz.
\label{eqn:limber}
\end{equation}
For the foreground large scale structure, 
$z_i=0,\ z_f=z_l,\ \Delta\chi(z)=\chi(z)$ where
$z_l$ is the redshift of the strong lens.  For the background large
scale structure, $z_i=z_l,\ z_f=z_s,\ \Delta\chi(z)=\chi(z_f)-\chi(z)$, 
where $z_s$ is the source redshift.  
We used the Peacock and Dodds (1996) formulation to obtain the
non-linear power from the linear transfer function given by
Bardeen et al (1986).
The comoving angular diameter distance is
\begin{equation}
D_A\equiv\chi(z)=c\int_0^z {\frac{dz}{H(z)}}
\label{eqn:chi}
\end{equation}
where H(z) is the Hubble constant at redshift z:
\begin{equation}
H(z)=H_0[(1+z)^2(\Omega_{\rm m} z+1)-\Omega_{\Lambda}z(z+2)]^{1/2}.
\end{equation}
For the comoving angular diameter distance $\chi$ we used the fitting formula from
Pen (1999). 
In a cosmological context, it is convenient to use comoving
angular diameter distances and conformal time, where light rays
propagate as they do in an empty universe (White \& Hu 2000).
The lensing weight is
\begin{equation}
w(z)=\frac{3}{2}\Omega_{\rm m} {H_0}^2g(z)(1+z)
\end{equation}
where
\begin{equation}
g(z)=\frac{[\chi(z_f)-\chi(z)][\chi(z)-\chi(z_i)]}{\chi(z_f)-\chi(z_i)}.
\end{equation}
In terms of our previous variables, for the background weak
lenses we have $D_{12}=\chi(z)-\chi(z_l)$, $D_{2s}=\chi(z_s)-\chi(z)$,
etc.   
The lensing weighting factor $g$ corresponds to the distance weighted
terms in Equation (\ref{eqn:sigma}).
A similar relation holds for the background lenses.

Table \ref{tab:rotation} gives the expected rotation angle for a
variety of quadruple lenses in the $\LCDM$ cosmology.  
Several simplifying assumptions were
made.  We attribute all the rotation to the furthest image.
The change in strong lens $\gamma$ between images is taken to be 0.3, which is
multiplied by the large scale structure shear in equation
(\ref{eqn:asym}).  What matters is not the change in the absolute
value of $\gamma$, but the change in each component.  We took the
variance of the large scale structure shear to be half of the
convergence, which corresponds to the variance in $\gamma_2$ in the
principal axis frame of the strong lens.  In practice, only
differences in rotation angles are observable, which depends on the
alignment angles of the shear at different image positions.  There is
a contribution to rotation from the structure in the foreground as
well as the background of the lens.  The variances were simply
added. The ratio of the foreground to background variance is listed in
column 5 in Table \ref{tab:rotation}. With all these caveats, we expect 
the expected rms image
rotation to be good to about a factor of two, which is comparable
to the expected errors on the theoretical lensing power spectrum.

\begin{table}
\begin{tabular}{l|ccccccc|}
lens & $z_{\rm lens}$ & $z_{\rm source}$& $\kappa_+$ & $\kappa_-$ & $\Delta
\phi$ (rms $^o$) & $(<z_{\rm lens})/(>z_{\rm lens})$ \\
\hline
B2045+265  &  0.87  &  1.28  & 0.018 & 0.003 & 0.4 & 35 \\
B0712+472   & 0.41  &  1.34  & 0.0076 &0.014 & 0.39 & 0.29\\
B1608+656  &  0.63  &  1.39  & 0.015 & 0.010 & 0.44 & 2.4\\
MG0414+0534 &   0.958 &   2.64  & 0.030  & 0.021 & 0.9 & 2.0\\
B1422+231  &  0.34  &  3.62 & 0.007& 0.049 & 1.2 & 0.02 \\
\hline
\end{tabular}
\caption{Expected image rotation angle for the most magnified image.
We assumed a characteristic change in strong lens $\gamma$ of 0.3
between images, and $\kappa=0.5$ at each image position.  Listed are
the lens and source redshifts, the expected rotation angle, and the
ratio of variances contributed from foreground compared to background
large scale structure. $\kappa_+$ is the standard deviation of
$\kappa$ and $\gamma$ due to foreground large scale structure, while
$\kappa_-$ is the contribution of the background.}  
\label{tab:rotation}
\end{table}

\section{Measuring rotation}

We only consider the simplest case.  The source is made up of three
components, $P^1,P^2,P^3$.  We define $P^3$ to sit at the origin, 
$P^2$ to sit at $(P^2_x,P^2_y)$, and $P^1$ at $(P^1_x,P^1_y)$,
We can always choose our coordinate system this way.

The lensed image appears at positions A, B, C, D also with three components
each.  We again define the position of the third component to be the origin,
and only consider relative distances. The apparent position of $A^2$
relative to $A^3$ is $ P^2={\bf D}_A A^2$, and similarly $P^1={\bf
  D}_A A^1$ where the deflection matrix ${\bf D}_A$ is the
inverse of the amplification matrix ${\bf A}$ defined in
equation (\ref{eqn:lens}) for image A; each image has its own deflection matrix.
It will be convenient to concatenate the two position column vectors
$P^1, P^2$ into a 2$\times$2 matrix ${\bf P}$.
We
can write similar equations for all images, resulting in an apparent
16 equations for 16 unknowns: ${\bf P}$ each has four unknowns, and
each symmetric amplification matrix has three unknowns.  The equations
are linear, and homogeneous: 
\beq
{\bf D}_A {\bf A}={\bf P}, \ \ {\bf D}_B{\bf B}={\bf P} , \
\ 
{\bf D}_C {\bf C}={\bf P}, \ \ {\bf D}_D{\bf D}={\bf P}.
\label{eqn:solve_rotate}
\eeq

It is clear that one could multiply each solution by a constant and
obtain a solution, so one must fix one more parameter.  Without loss
of generality, one could fix $P^2_x=1$, which just fixes a length
scale.  If none of the amplifications are known, the solutions are
clearly degenerate between a small image that is strongly magnified
and a larger image that is less magnified.  A further degeneracy
occurs because we can multiply each equation on the left by an
arbitrary shear matrix.  Since we do not know the intrinsic
location of the source substructure positions, this is indistinguishable
from a constant shear applied to both the lens and the source.  This
corresponds to a shear plane between strong lens and source.  For a
shear between observer and strong lens, one can simularly symmetrize
the deflection matrix by multiplying both the lens and the sources
by the inverse shear matrix.  This is in accordance with Schneider
(1997).  Thus, we only measure the differences in shears at the
image positions.

Then we have 16 equations
for 15 unknowns, which is over-determined by one.  This allows us to
solve for one rotation angle.  If one assumes the rotation to be
dominated by the most magnified image, say $A$, we simply allow ${\bf
  D}_A$ to be non-symmetric.  This allows one to solve for the rotation
angle. 

In general, however, the large scale structure shear has two
independent components $\gamma_1,\gamma_2$.  To make progress, one can
assume them to be Gaussian distributed with standard deviation
$\sigma$.  Their one point function is
independent,
$P(\gamma_1,\gamma_2)=N(\gamma_1,\sigma)N(\gamma_2,\sigma)$.
Equation
(\ref{eqn:asym}) expresses the off-diagonal components of each
deflection matrix in terms of the two large-scale structure shear components.  If we fix
a value of $\gamma_1$, we can solve Equation (\ref{eqn:solve_rotate}) for
$\gamma_2$, 
giving us an implicit definition of $\gamma_2(\gamma_1)$.
Then integrating over all possible values of $\gamma_1$ weighted by
the probability gives the total likelihood for an assumed $\sigma$:
\beq
L(\sigma) = \int P(\gamma_1,\gamma_2(\gamma_1))\, d\gamma_1.
\eeq
One can then solve for the maximum likelihood value of $\sigma$.

\section{Discussion}

If one wishes to observe this effect, several challenges must be
overcome.  Sources must have at least three localizable components,
and the position of each must be measured very precisely, to better
than 1\% of the component separation.  Positions are also affected by
changes in the lens from one component position to the next.  The
latter effect is expected to be significantly smaller than rotation,
because it depends on the change in shear on small scales, while the
rotation depends on the large scale structure shear itself.
Basically, the rotation is of order of the large scale structure shear.
The variation of the large scale structure shear on source substructure scales
is smaller than the shear itself.

We quantify this as follows: let us assume that the components of
the lens have a separation of 10 mas.  The model assumed that the
shear was constant over the apparent size of the source.  The
change in shear across the lens is given by 
$\xi_\kappa(\Delta \theta=10 {\rm mas})$.  From Figure \ref{fig:corr}
we see that $\kappa$ has a differential variance of around $10^{-4}$
at this separation, corresponding to a percent change in lensed
length scales.  This is of comparable magnitude to the rotation
effect, which makes it desirable to have more than three components
to check.  In practice, the actual lensing substructure is suppressed
by several factors.  Figure \ref{fig:corr} shows the weak lensing correlation
in the absence of a strong lens.  Lines converge behind the lens,
which reduces the corresponding physical scales, and thus the total
variance.  And only the variation in shear principle axes across
three sub-image positions affects the rotation, which is suppressed
by another dimensionless factor.

It is worth noting that HS proved that the
rotation $\omega$ is identical to the B-mode of the shear, up to a
constant.  In this calculation, we only considered the local value of
shears, while the E-B decomposition is non-local.  We therefore should
consider the 4 entries in the magnification matrix to be independent.
In principle, if one had a very high number density of background
sources, for example from reionization (Pen 2004), one could also
solve for the strong lensing E-B decomposed map (Pen 2000).  Such
a decomposition also allows one to distinguish between structures
in the strong lens plane, and contributions from large scale structure
along the line of sight.  The signal to noise required for such an
exercise is difficult to achieve with current technology.  Using
quadruply imaged sources allows one to reduce the pairwise noise,
and thus solve for the rotational component.

We also note that our calculation did not use the Born
approximation.  We modeled each lens plane by a quadratic
potential as inferred from the $\kappa$ correlation function, 
and computed the full deflection trajectories.  Recently,
calculations of rotations by large scale structure at different lens
planes resulted in different answers, depending on whether
calculations were done with light rays (HS) or
with shears on light bundles (CH).  In the CH calculation, ray
bundles were propagated on unperturbed trajectories, but distortions
were accumulated.  The distortions are relative deflections of
neighboring rays, which accumulate rotations through the same
effect as discussed in this paper.  This allows CH to see
rotation without considering the explicit deflection of the centers
of each ray.  In terms of a full Taylor expansion of light rays
along unperturbed paths, this includes some of the effects beyond
the Born approximation.  HS included all leading order corrections
to the Born approximation, and thus differed in their answer.
In our case, the
angular scales involved are very small, and we can assume the
rotation to be dominated by the common large scale structure shear at
different image positions.  The CH and HS approaches do not differ in
our model, since the potential is taken to be quadratic, and all
second derivatives are constant.

Our result differ from that of Schneider (1997) in the nature
of the effect.  Schneider (1997) constructs
an equivalent single plane strong lens from a strong lens plus
constant weak shear.  Our calculation circumvents the assumption
that the weak lens screen is spatially constant.

Our model assumes a source plane consisting of three compact components
with unknown positions.  If all three components are quadruply lensed,
there are a total of 12 angular positions, i.e., we have 24 observables.
We subtract 8 degrees of freedom for the global lens model, because we
do not know the macroscopic lens deflection angle. This leaves 16 
constraints.  Then using Equation (\ref{eqn:solve_rotate}) we can solve for 
the relative positions
in the source plane modulo a scaling (due to the mass sheet degeneracy),
which is three numbers,
as well as the values of the amplification matrix at each image
position, which is four sets of 3 numbers, plus one rotation, for a total
of 16 parameters. This suggests that small scale structure in a single lens is
observationally distinguishable from weak shear at a different redshift.  

We have made several approximations, which will affect the results
at some level.  We used the difference in shears at the two furthest
image positions.  In the analysis, we then assumed that three of the
images had no rotation, and only the furthest accounted for the
multi-plane rotation.  Reality is more complex, and all images
have differential weak lensing shear, and thus some level of differential
rotation.  Simulations are
needed to quantify this simplifying assumption.

In this paper, we have concentrated on the rotation induced
by the weak large-scale structure. However, for the cases where
one has multiple lenses along the line of sight, the rotation can
be more significant. For the best-fit model in Chae, Mao \& Augusto (2001), we
find that the two Jacobian matrices are given by
\beq
J=
\left[ 
\begin{array}{cc}
0.176 & -0.093 \\ 
-0.258 & -0.595
\end{array}
\right ], ~~~~
\left [
\begin{array}{cc}
0.888 & -0.174 \\ 
-0.148 & 0.481
\end{array}
\right ],
\eeq
for images A and D, corresponding to magnifications $-7.78$ and $2.49$
respectively. Clearly for the more highly magnified image $A$,
the Jacobian is highly asymmetric and the rotation is quite significant.

\section{Conclusions}

We have computed the expected rotation from uncorrelated foreground
and background large scale structure in strong lensing systems.  We
have shown that this effect is in principle observable with precise
VLBI imaging of quadruply imaged lens systems.  
The rotation is physical and observable in the example of a source
plane consisting of three point sources.
If observed, it can unambiguously determine if the flux
anomaly problem is caused by substructure on the lens plane, or by
uncorrelated structures along the line of sight.

In addition, it demonstrates how one can extract information about
small scale dark matter structure along the
line of sight to a lens.  A measurement of rotation would 
measure the variance of $\kappa$ from large scale structure at the
smallest scales.  This affects the scatter in supernovae
lensing effects, and may have potential to measure the primordial
power on small scales, and possibly a tilt or running spectral
index.

We thank Peter Schneider for pointing out a flaw in an earlier
version of the paper, T. York and K.-H. Chae for helpful discussions, 
and Pengjie Zhang for the lensing Limber code.

{}
\end{document}